\def\ref{\par\noindent\hang}

\def\spose#1{\hbox to 0pt{#1\hss}}
\def\approxlt{\mathrel{\spose{\lower 3pt\hbox{$\sim$}}
        \raise 2.0pt\hbox{$<$}}}
\def\approxgt{\mathrel{\spose{\lower 3pt\hbox{$\sim$}}
        \raise 2.0pt\hbox{$>$}}}

\def\multleft#1{\hbox to size{\vbox {\halign {\lft{##}\cr #1}}\hfill}\par}
\def\multright#1{\hbox to size{\vbox {\halign {\rt{##}\cr #1}}\hfill}\par}

\def\degmark{^\circ}
\def\boxit#1{\vbox{\hrule\hbox{\vrule\kern3pt\vbox{\kern3pt
          #1 \kern3pt}\kern3pt\vrule}\hrule}}

\def\erg{{\rm\thinspace erg}}

\def\keV{{\rm\thinspace keV}}
\def\km{{\rm\thinspace km}}

\def\Mpc{{\rm\thinspace Mpc}}
\def\Msun{\hbox{$\rm\thinspace M_{\odot}$}}

\def\s{{\rm\thinspace s}}

\def\Hz{{\rm\thinspace Hz}}

\def\ergps{\hbox{$\erg\s^{-1}\,$}}

\def\kmps{\hbox{$\km\s^{-1}\,$}}

\def\kmpspMpc{\hbox{$\kmps\Mpc^{-1}$}}

\documentclass{article}
\usepackage{psfig}
\usepackage{emulateapj}
\usepackage{times}
\usepackage{amssym}

\makeatletter
\@ifundefined{chapter}{\def\thebibliography#1{
\section*{References}
  \list
  {\relax}{\setlength{\labelsep}{0em}
        \setlength{\itemindent}{-\bibhang}
        \setlength{\itemsep}{\parskip}
        \setlength{\parsep}{0pt}
        \setlength{\leftmargin}{\bibhang}}
    \def\newblock{\hskip .11em plus .33em minus .07em}
    \sloppy\clubpenalty4000\widowpenalty4000
    \sfcode`\.=1000\relax}}%
{\def\thebibliography#1{
  \list
  {\relax}{\setlength{\labelsep}{0em}
        \setlength{\itemindent}{-\bibhang}
        \setlength{\itemsep}{\parskip}
        \setlength{\parsep}{0pt}
        \setlength{\leftmargin}{\bibhang}}
    \def\newblock{\hskip .11em plus .33em minus .07em}
    \sloppy\clubpenalty4000\widowpenalty4000
    \sfcode`\.=1000\relax}}
 
\newlength{\bibhang}
\let\@internalcite\cite
\def\cite{\@ifstar{\citey}{\citefull}}
\def\citefull{\def\astroncite##1##2{##1\ ##2}\@internalcite}
\def\citey{\def\astroncite##1##2{##1\ (##2)}\@internalcite}
\def\citeyear{\def\astroncite##1##2{##2}\@internalcite}
\def\citename{\def\astroncite##1##2{##1}\@internalcite}
\def\@citex[#1]#2{\if@filesw\immediate\write\@auxout{\string\citation{#2}}\fi
  \def\@citea{}\@cite{\@for\@citeb:=#2\do
    {\@citea\def\@citea{; }\@ifundefined
       {b@\@citeb}{{\bf ??}\@warning
       {Citation `\@citeb' on page \thepage \space undefined}}%
{\csname b@\@citeb\endcsname}}}{#1}}
 
\def\@cite#1#2{#1\if@tempswa #2\fi} 
\def\@biblabel#1{}
 
\def\astroncite#1#2{#1\ #2}
\makeatother

\begin{document}

\title{On the lack of X-ray iron line reverberation in MCG--6-30-15:\\
Implications for the black hole mass and accretion disk structure}

\author{Christopher~S.~Reynolds\altaffilmark{1,2}}
\altaffiltext{1}{JILA, Campus Box 440, University of Colorado, Boulder
CO~80303, USA}
\altaffiltext{2}{Hubble Fellow}

\begin{abstract}
We use the method of Press, Rybicki \& Hewitt (1992) to search for time
lags and time leads between different energy bands of the RXTE data for
MCG--6-30-15.  We tailor our search in order to probe any reverberation
signatures of the fluorescent iron K$\alpha$ line that is thought to
arise from the inner regions of the black hole accretion disk.  In
essence, an optimal reconstruction algorithm is applied to the continuum
band (2--4\,keV) light curve which smoothes out noise and interpolates
across the data gaps.  The reconstructed continuum band light curve can
then be folded through trial transfer functions in an attempt to find
lags or leads between the continuum band and the iron line band
(5--7\,keV).  We find reduced fractional variability in the line band.
The spectral analysis of Lee et al. (1999) reveals this to be due to a
combination of an apparently constant iron line flux (at least on
timescales of ${\rm few}\times 10^4\s$), and flux correlated changes in
the photon index.  We also find no evidence for iron line reverberation
and exclude reverberation delays in the range 0.5--50\,ksec.  This
extends the conclusions of Lee et al. and suggests that the iron line
flux remains constant on timescales as short as 0.5\,ksec.  The large
black hole mass ($>10^8\Msun$) naively suggested by the constancy of the
iron line flux is rejected on other grounds.  We suggest that the black
hole in MCG--6-30-15 has a mass of $M_{\rm BH}\sim 10^6-10^7\Msun$ and
that changes in the ionization state of the disk may produce the
puzzling spectral variability.  Finally, it is found that the 8--15\,keV
band lags the 2--4\,keV band by 50--100\,s.  This result is used to
place constraints on the size and geometry of the Comptonizing medium
responsible for the hard X-ray power-law in this AGN.
\end{abstract}

\begin{keywords}
{galaxies:Seyfert, galaxies:individual:MCG--6-30-15, line:formation,
methods:statistical, X-ray:galaxies}
\end{keywords}

\section{Introduction}

The X-rays from active galactic nuclei (AGN) are thought to originate from
the innermost regions of an accretion disk around a central supermassive
black hole.  Thus, in principle, the study of these X-rays should allow one
to probe the immediate environment of the accreting black hole as well as
the exotic physics, including strong-field general relativity, that
operates in this environment.  

In the past decade X-ray astronomy has begun to fulfill that promise.  Both
{\it EXOSAT} and {\it Ginga} discovered iron K-shell features (including
the K$\alpha$ fluorescent line of cold iron at 6.4\,keV) in the X-ray
spectra of Seyfert galaxies which were interpreted as `reflection' of the
primary X-ray continuum by cold, optically-thick material in the immediate
vicinity of the black hole (Guilbert \& Rees 1988; Lightman \& White 1988;
Nandra et al. 1989; Nandra, Pounds \& Stewart 1990; Matsuoka et al. 1990).
It was suggested that this cold reflecting material was the putative
accretion disk of AGN models.  With the launch of {\it ASCA} and the advent
of medium resolution spectroscopy, the iron line in several objects was
shown to be broad ($\sim 80\,000\kmps$ FWZI) and skewed (Tanaka et
al. 1995; Nandra et al. 1997).  The overall line profiles are in good
agreement with models for fluorescent line emission from the innermost
regions of geometrically-thin black hole accretion disks (Fabian et
al. 1989).  Such data allow us to address issues such as the location of
the radius of marginal stability, the spin of the black hole, and the
inclination distribution of various classes of AGN (see Reynolds 1999 and
references therein for a review of these studies).  In the current, {\it
RXTE} era, we can now probe the iron line and Compton reflection hump in
individual objects in some detail (e.g., MCG-5-23-16, Weaver et al. 1998;
MCG--6-30-15, Lee et al. 1998, 1999a; NGC~5548, Chiang et al. 1999).

While these spectral studies have been successful, a complete picture of
the AGN phenomenon is not possible without addressing the timing
properties.  Timing studies are important for two intertwined reasons.
Firstly, AGN are {\it inherently} variable systems.  In general, the
variability timescale in a given object is seen to shorten as one considers
higher frequency radiation.  In the X-ray and $\gamma$-ray bands, dramatic
variability has been seen in many Seyfert galaxies with doubling timescales
of only a few minutes (e.g. see Reynolds et al. 1995).  Although it is
poorly understood to date, the nature of this violent variability is a
vital component of any final AGN model.  Careful characterization of the
timing properties, as well as determining the observed spectral evolution
during dramatic temporal events, is required if we are to understand this
phenomenon.  

Secondly, timing studies are needed to break certain degeneracies that
exist in models which, to date, have only been constrained by purely
spectral data.  The spin of the black hole in MCG--6-30-15 provides an
excellent example of such a degeneracy --- by fitting the `very-broad'
state (Iwasawa et al. 1996) of the iron line in this object with models
consisting of a thin, disk-hugging corona, Dabrowski et al. (1997) inferred
that the black hole in this AGN must be almost maximally rotating, with a
dimensionless spin parameter of $a>0.94$.  However, by including line
emission from within the radius of marginal stability, Reynolds \&
Begelman (1997) showed that a geometry in which the X-ray source is at some
height above the disk plane can produce the same line profile even if the
black hole is completely non-rotating.  While there are subtle spectral
differences between the two scenarios (Young, Ross \& Fabian 1998) the most
obvious way of distinguishing these scenarios is through their timing
properties.  The Reynolds \& Begelman (1997) geometry predicts substantial
time delays between fluctuations in the primary power-law continuum and the
responding fluctuations in the iron line.  More generally, the
reverberation characteristics of the iron line contain tremendous
information on the mass and spin of the black hole as well as the geometry
of the X-ray source (Stella 1990; Reynolds et al. 1999).

The observational situation is more complex.  Lee et al. (1999b) and
Chiang et al. (1999) have analyzed extensive {\it RXTE} datasets for
MCG--6-30-15 and NGC~5548, respectively, in order to study the timing
properties and spectral variability.  In both of these objects, the same
pattern of spectral variability is seen.  Firstly, the X-ray photon
index displays flux-correlated changes in the sense that the source is
softer when it is brighter.  Secondly, and more surprisingly, the iron
line flux was found to be constant over the timescales probed by these
direct spectral studies ($\sim 50-500$\,ksec).  As discussed by both
sets of authors, these results are difficult to interpret in the
framework of standard X-ray reflection models since the breadth of these
lines indicate that they originate from a small region.  It appears that
some feedback mechanism regulates the amount of iron line emission in
order to produce approximately constant iron line flux.  Flux-correlated
changes in the ionization state of the disk represent one such mechanism
(we discuss this in more detail in Section 5 of this paper).  Unless
this feedback mechanism operates instantaneously, we might still expect
variability of the iron line flux on short timescales.

Driven by these motivations, this paper addresses the problem of
determining causal relationships between light curves in different X-ray
bands, with particular emphasis on timescales shorter than those that
can be probed by direct spectroscopy.  In particular, we use the long
{\it RXTE} observation of the bright Seyfert 1 galaxy MCG--6-30-15
reported by Lee et al. (1999a,b) and consider the relationship between
the 2--4\,keV band (hereafter called the continuum band) and the
5--7\,keV band which contains most of the iron line photons (and
hereafter called the line band).  An important special case is one in
which there is a linear transfer function relating one band to the
other:
\begin{equation}
b(t)=\int_{-\infty}^{\infty}d\tau\, \Psi(\tau) a(t-\tau),
\end{equation}
where $a(t)$ and $b(t)$ are continuum and line band fluxes respectively,
and $\Psi$ is the transfer function.  Such relationships between bands
contain much of the important physical information, such as the
reverberation characteristics of the iron line.

Mathematically, the linear transfer equation can be easily inverted using
Fourier methods to obtain,
\begin{equation}
\Psi(t)=\frac{1}{2\pi}\int^{\infty}_{-\infty}d\omega\,e^{-i\omega t}\frac{\tilde
{a}(\omega)}{\tilde{b}(\omega)},
\end{equation}
where $\tilde{a}(\omega)$ represents the Fourier transform of $a(t)$.
However, in real situations, a large number of regularly sampled
measurements are required to obtain an accurate deconvolution using this
simple method.  More often, deconvolution is achieved using maximum entropy
techniques or some other regularization method (Horne et al. 1991; Krolik
et al. 1991).

Another common approach (and one that is often used with less well sampled
data) is to compute cross-correlation functions (CCFs), or some variant
thereof which accounts for the finite and irregular sampling often
encountered in real data.  The discrete correlation function (DCF; Edelson
\& Krolik 1988) is one example of such a variant.  Lee et al. (1999b) apply
such methods to the observation of MCG--6-30-15 considered in this paper
and detect both phase and time lags between {\it RXTE} bands (also see
Nowak \& Chiang 1999).  While these methods are powerful, it can be
difficult to separate subtle time leads/lags from the autocorrelation
properties of the data.

Here, we take an alternative approach which is heavily based on the method
of Press, Rybicki \& Hewitt (1992; hereafter PRH92).  In essence, we use
the correlation properties of the continuum band data to reconstruct an
optimal continuum light curve in which the data gaps have been
interpolated.  Most importantly, we also compute the expected deviation of
the continuum flux from the interpolated curve.  The reconstructed
continuum band light curve is convolved with a trial transfer function and
compared with the line band light curve in a $\chi^2$ sense.  We then
examine changes in the $\chi^2$ statistic as a function of the parameters
that define the trial transfer function.

Section~2 recaps the PRH92 method.  This is then applied to the {\it RXTE}
data for MCG--6-30-15 in Section~3.  The robustness and validity of our
approach is demonstrated by applying this method to simulated data
(Section~4).  Section~5 draws together our results and discusses their
implications for the nature of this source.  In particular, we argue that
the black hole in this AGN has a mass of only $10^6-10^7\Msun$.  In order
to explain the spectral variability, it is suggested that there are flux
correlated changes in the ionization state of the surface layers of the
accretion disk.   Section~6 presents a short summary of the results and
relevant astrophysical implications.

\section{The problem and method of solution}

\subsection{The optimal reconstruction}

The continuum band light curve is reconstructed from the data using the
technique of PRH92.  For completeness, this section summarizes their
method.  The reader who is primarily interested in the application of this
method may skip to Section~3.   

Suppose that the true flux of the source at time $t$ is $s(t)$, but we
measure $y(t)=s(t)+n(t)$, where $n(t)$ is the noise in the measurement.  In
our case, the noise is Poisson in nature.  Our knowledge of $s(t)$ is
further impeded by the fact that the measurement is only made at a finite
number of times $t_i$, where $i=1,...,N$.  We denote $y(t_i)$ as $y_i$ and
refer to this as the continuum data vector.

We seek an optimal reconstruction of $s(t)$ which is continuous in time,
$\hat{s}(t)$, such that
\begin{equation}
\langle e^2(t)\rangle\equiv \langle[\hat{s}(t)-s(t)]^2\rangle
\end{equation}
is minimized for all $t$.  As usual, angle brackets denote the expectation
value.  We impose that $\hat{s}(t)$ is linear in the data vector in the
sense that
\begin{equation}
\hat{s}(t)=\sum^{N}_{i=1}q_i(t)y_i,
\end{equation}
where $q_i(t)$ are a set of inverse response functions that are also
continuous in time.   

Assuming that the noise is uncorrelated with both $s(t)$ and itself, PRH92
showed that eqn (3) can be minimized to yield,
\begin{equation}
\hat{s}(t)=\sum^{N}_{i,j=1}\langle s(t_i)s(t)\rangle(B^{-1})_{ij}\,y_j.
\end{equation}
Here, 
\begin{equation}
B_{ij}=\langle s(t_i)s(t_j)\rangle+\langle n_i^2\rangle \delta_{ij}
\end{equation} 
is the total covariance matrix.  To keep the notation concise, PRH92 define
the correlation statistics:
\begin{eqnarray}
c_i&\equiv &\langle s(t_i)s(t)\rangle\\
C_{ij}&\equiv & \langle s(t_i)s(t_j)\rangle\\
{\cal C}(t)&\equiv & \langle s(t)s(t)\rangle
\end{eqnarray}
These functions define what PRH92 call the `covariance model'.  The
expected variance of the real signal from the optimal reconstruct in eqn
(5) is then given by
\begin{equation}
\langle e^2(t)\rangle={\cal C}(t)-\sum^{N}_{i,j=1}c_i(t)(B^{-1})_{ij}c_j(t).
\end{equation}

Once the covariance model is known, eqns (5) and (10) define the optimal
reconstruction of the continuum light curve together with a statistic
measuring the expected deviation of the real signal from the
reconstruction.

\subsection{The covariance model}

Here, again, we follow the method of PRH92 to determine the covariance
model for our continuum data.   At this stage, we make the assumption that
the underlying process is statistically stationary so that
\begin{equation}
\langle s(t_1)s(t_2)\rangle=\langle s(0)s(t_2-t_1)\rangle=C(t_2-t_1),
\end{equation}
where $C(\tau)$ is the autocorrelation function that we have to determine.
This function is related to the first order structure function $V(\tau)$ by
\begin{equation}
C(\tau)=\langle s^2 \rangle - V(\tau),
\end{equation}
and $V(\tau)$ can be approximated by forming pair-wise estimates for all
distinct pairs of data points in the continuum light curve, and then binning
by the time lag of the pairs.  We find that the analytic form
\begin{equation}
V_{\rm anal}(\tau)=A\left(\frac{(\tau/\tau_0)^{\alpha_1}}{1+(\tau/\tau_0)^{\alpha_1}}\right)^{\alpha_2}
\end{equation}
fits the structure functions of this paper well.   In fact, the
reconstruction is fairly insensitive to the exact analytic form used to
approximate the structure function.

Making the reasonable assumption that $C(\tau)\rightarrow \langle
s\rangle^2$ as $\tau\rightarrow\infty$, our final expression for the
autocorrelation function is
\begin{equation}
C(\tau)=\langle s\rangle^2 + V_{\rm anal}(\infty)-V_{\rm anal}(\tau).
\end{equation}

\vspace{0.5cm}

\section{Application to MCG--6-30-15}

In this Section, we apply the method outlined above to a long {\it RXTE}
observation of the bright Seyfert 1 galaxy MCG--6-30-15.

\subsection{The RXTE data}

\begin{figure*}
\centerline{
\psfig{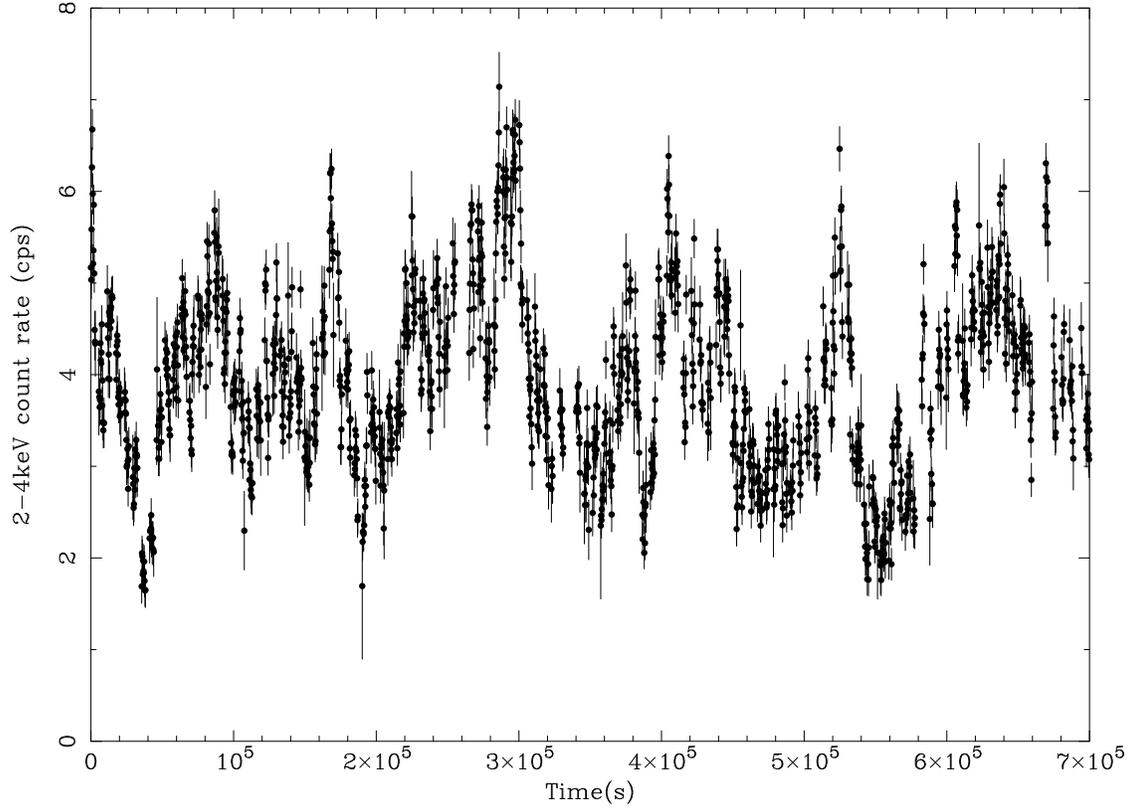}
}
\caption{2--4\,keV band, 3-PCU, light curve for the 1997-Aug-4 {\it RXTE}
observation of MCG--6-30-15.   For display purposes, a bin size of 256\,s
has been used, although 64\,s bins are used in the analysis presented in
this paper.}
\end{figure*}

{\it RXTE} observed MCG--6-30-15 for approximately $7\times 10^5\s$
starting on 4-Aug-1997.  We retrieved these data from the NASA-{\sc
heasarc} public archive situated at the Goddard Space Flight Center.  Our
data reduction closely parallels that of Lee et al. (1999a) who has studied
the spectral characteristics of this observation.  Since, as mentioned in
the introduction, we are interested in the soft X-ray continuum and the
iron line band, the Proportional Counter Array (PCA) is the appropriate
instrument for us to consider.  Examining the housekeeping files for this
observation reveals that Proportional Counter Units (PCUs) 3 and 4 suffer
occasional breakdown and shut off.  Hence, we do not consider data from
these units and, instead, extracted {\sc standard-2} data from PCUs 0--2.
We applied fairly standard faint-source screening criteria to these data:
the source must be at least $10\degmark$ above the Earth's limb ({\sc
elv}$>$10), the source must be located within $0.02\degmark$ of the nominal
pointing position ({\sc offset}$<$0.02), there must be at least three PCUs
on ({\sc num\_pcu\_on}$>$2), it has been at least 30 minutes since a
passage of the South Atlantic Anomaly ({\sc time\_since\_saa}$>$30), and
the electron background is not too high ({\sc electron0}$<$0.1).  After
application of these screening criteria, approximately $3.5\times 10^5\s$
of `good' data remain.  From these data, 2--4\,keV and 5--7\,keV light
curves were extracted using 64\,s bins.   We also extracted the 8--15\,keV
light curve which we will use in Section 3.3.

The background was estimated using the L7-240 background models which are
appropriate for faint sources such as AGN.   Background light curves were
computed and subtracted from the measured light curves in order to form the
final background subtracted light curves that we shall use in our study.
Figure~1 shows the continuum band (2--4\,keV) light curves that results
from this procedure.   For clarity, the light curve shown in this figure has
been binned with 256\,s bins.

\begin{figure*}
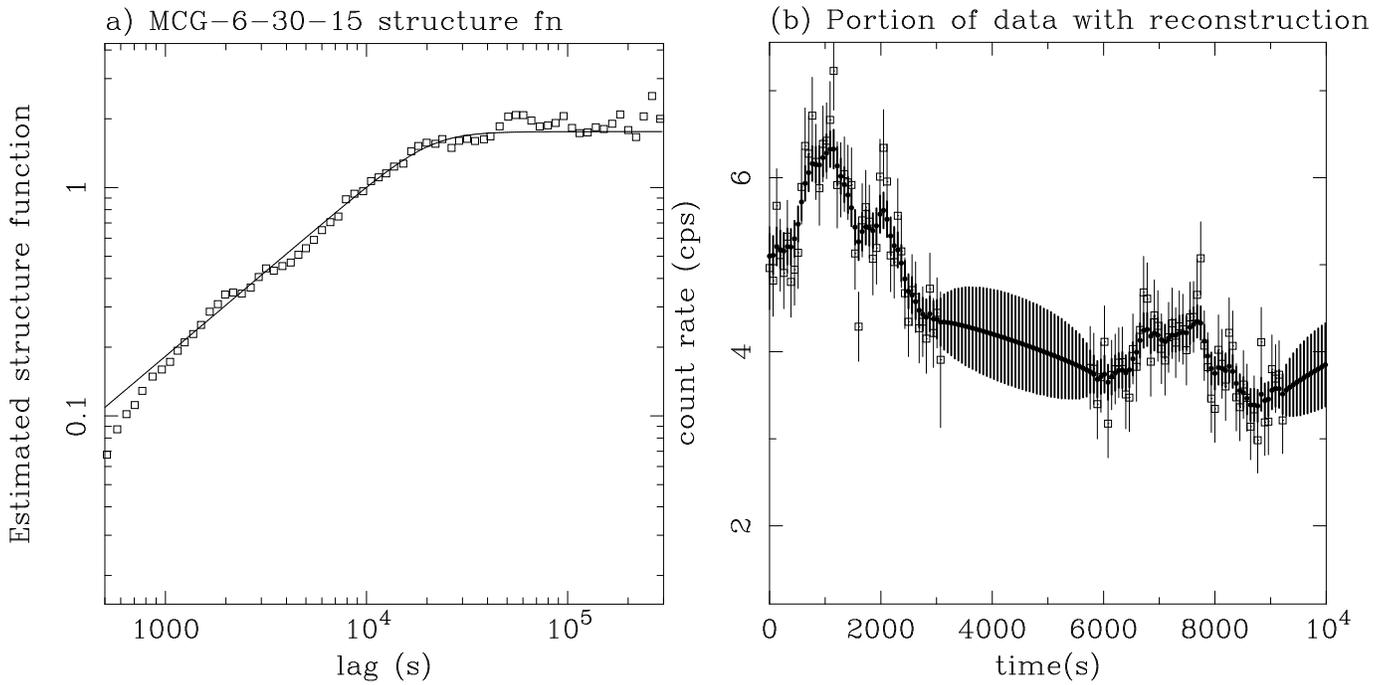

\hbox{
\psfig{figure=mcg6_sf.ps,width=0.47\textwidth,angle=270}
\psfig{figure=mcg6_reconstruction.ps,width=0.5\textwidth,angle=270}
}
\caption{Panel (a) shows the structure function for the 2--4\,keV band data
from MCG--6-30-15 (squares), together with our analytic approximation
(solid line).  Panel (b) shows a portion of the 2-4\,keV light curve (open
squares and thin error bars) together with the reconstructed light curve
(filled circles and heavy error bars).}
\label{mcg6recon}
\end{figure*}

\subsection{Searching for lags and leads}

We now apply the procedure outlined in Section~2 to these lightcurves.  To
begin with, we must estimate the structure function for these data.
Figure~2a shows a pair-wise estimate of the continuum band structure
function obtained following the method of PRH92.  This figure also shows
our analytic approximation which is given by eqns (13) and (14) with
\begin{eqnarray}
A&=&1.757\,{\rm cps}\\
\tau_0&=&2.117\times 10^3\s\\
\alpha_1&=&3.805\\
\alpha_2&=&0.195\\
\langle s\rangle&=&3.91\,{\rm cps}
\end{eqnarray}
Using this covariance model, the PRH92 reconstruction was applied to the
continuum light curve using $N=5000$ data points.  A portion of the
resulting reconstructed light curve is shown in Fig.~2b.

The next stage in the procedure is to convolve the reconstructed
continuum band light curve with a trial transfer function and compare
the result with the line band light curve in a $\chi^2$ sense.  We can
then minimize the $\chi^2$ statistic in order to constrain free
parameters in the trial transfer function.  We also minimize $\chi^2$
over multiplicative and additive offsets between the continuum and line
band light curves, i.e. we set
\begin{equation}
b(t)=B\int_{-\infty}^{+\infty}\psi(\tau) a(t-\tau)\,d\tau + K,
\end{equation}
and minimize over $B$ and $K$ as well as the parameters describing the
trial transfer function $\psi$.  

In this work, we choose two trial transfer functions.  The first
represents the case where some fraction $f_{\rm tr}$ of the line band
flux is a delayed copy of the continuum band with a time delay $t_{\rm
tr}$:
\begin{equation}
\psi_1(t)= (1-f_{\rm tr})\delta(t)+f_{\rm tr}\delta(t-t_{\rm tr}).
\end{equation}
The second represents the case where some fraction $f_{\rm tr}$ of the line
band flux is a delayed and smeared copy of the continuum band flux, where a
Gaussian kernel is used:
\begin{equation}
\psi_2(t)= (1-f_{\rm tr})\delta(t)+\frac{f_{\rm tr}}{\sigma_{\rm tr}}\sqrt{\frac{2}{\pi}}\exp\left( -\frac{(t-t_{\rm tr})^2}{2\sigma_{\rm tr}^2}\right),
\end{equation}

No extrapolations were performed during this procedure.  In order to
avoid extrapolating, the $\chi^2$ statistic was calculated using a
subset of data points.  For the trial transfer function $\psi_1$, only
data during times $t_{\rm start}+t_{\rm tr,max}<t<t_{\rm end}-t_{\rm
tr,max}$ were used to compute $\chi^2$, where $t_{\rm start}$ and
$t_{\rm end}$ are the times of the start and end of the reconstructed
continuum light curve.  For $\psi_2$, $\chi^2$ is computed based upon
data from times $t_{\rm start}+t_{\rm tr,max}+2\sigma_{\rm
tr,max}<t<t_{\rm end}-t_{\rm tr,max}-2\sigma_{\rm tr,max}$.

Figure~3 shows the $\chi^2$ surfaces and confidence contours once this
procedure has been performed.  When displaying the $\chi^2$ surfaces, we
plot $\log_{10}(\chi^2-\chi^2_{\rm min}+1)$ in order to highlight the
topography of the surface near the global minimum in the surface.  It
can be seen that the minimum of the $\chi$2 surface corresponds to the
two lines $f_{\rm tr}=0$ and $t_{\rm tr}=0$, i.e. no time delayed
component of the line band light curve is detected.  Here we only show
the results for $\psi_1$ --- the $\psi_2$ results are trivial
(i.e. $\chi^2$ surface is completely flat) since the preferred solution
always have $f_{\rm tr}=0$.  The best fit values of the multiplicative
and additive constants are $B=0.78$ and $K=0.90$.

\begin{figure*}
\hbox{
\psfig{figure=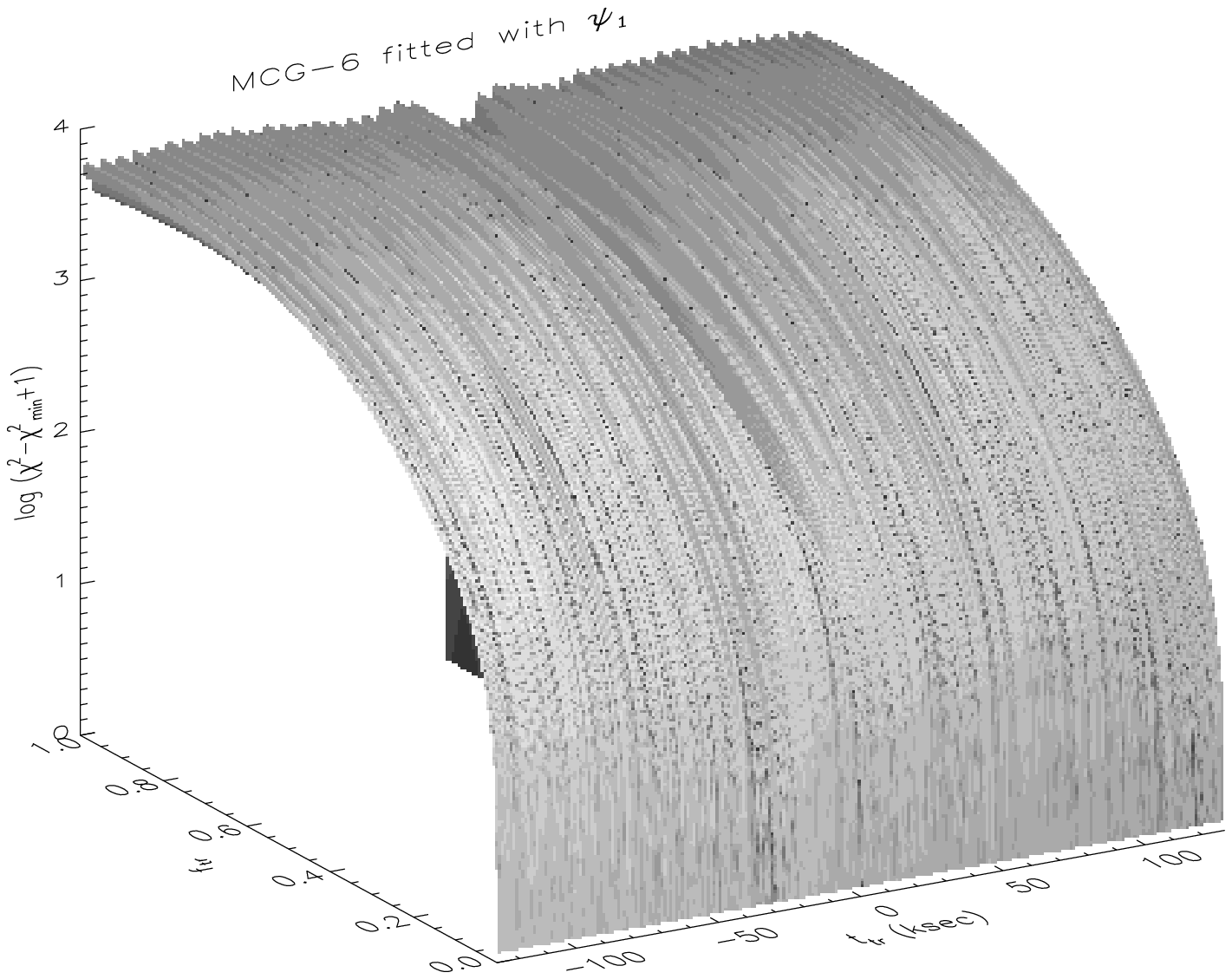,width=0.5\textwidth}
\psfig{figure=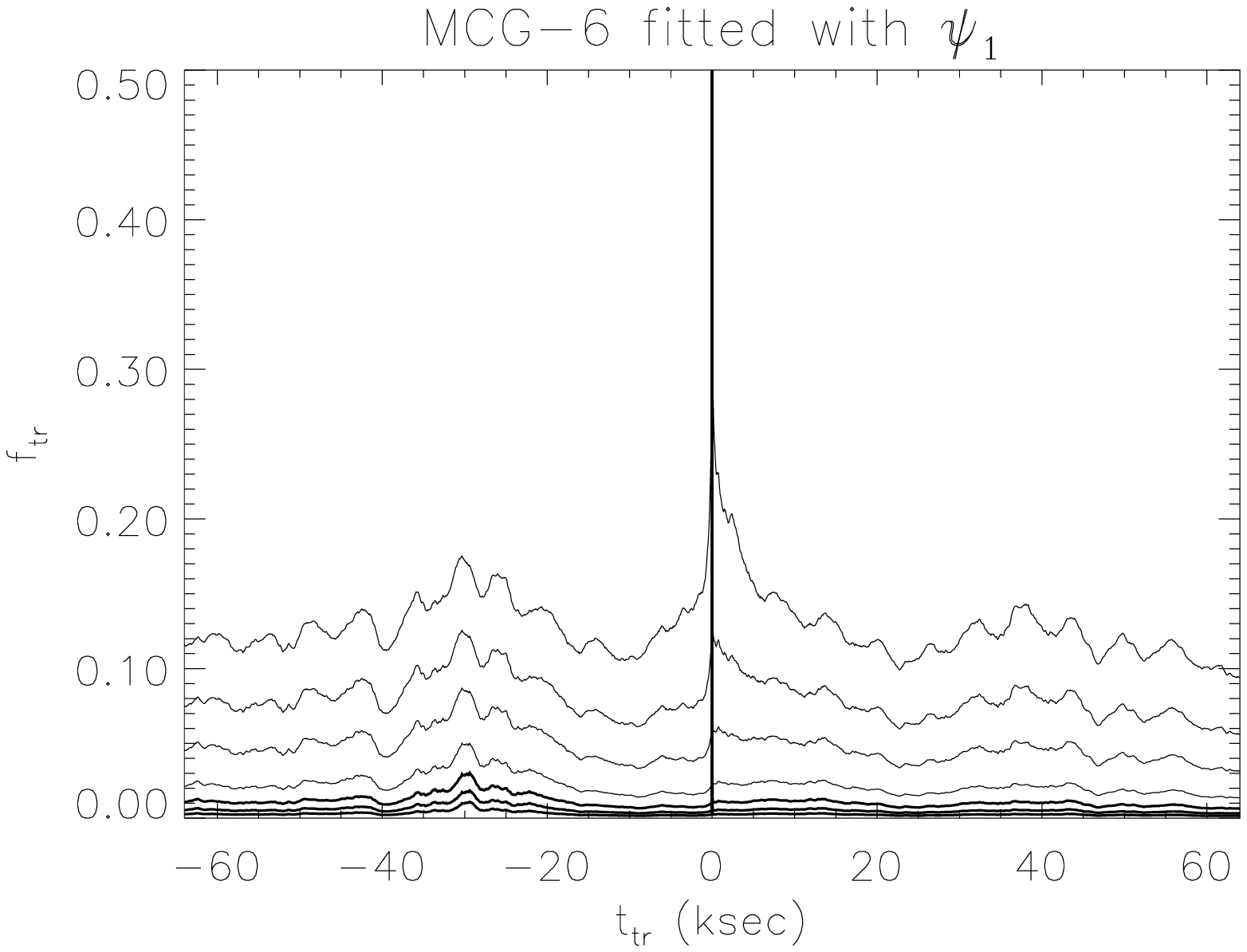,width=0.5\textwidth}
}
\caption{Results for MCG--6-30-15:  $\chi^2$ surfaces and confidence 
contours resulting from applying trial transfer function $\psi_1$ to the
reconstructed continuum light curves and comparing with the line band
light curve.  Surfaces are plotted using $log_{10}(\chi^2-\chi^2_{\rm
min}+1)$ as the ordinate in order to display the topography of the
region near the minimum.  Contours are shown the following levels:
$\chi^2-\chi^2_{\rm min}=2.3,4.6,9.2,20,50,100,200$.  The first three of
these contours correspond to $1\sigma$, 90\% and 95\% for two
interesting parameters and are shown in bold.}
\end{figure*}

\subsection{The overall time delays between bands}

\begin{figure*}
\centerline{
\psfig{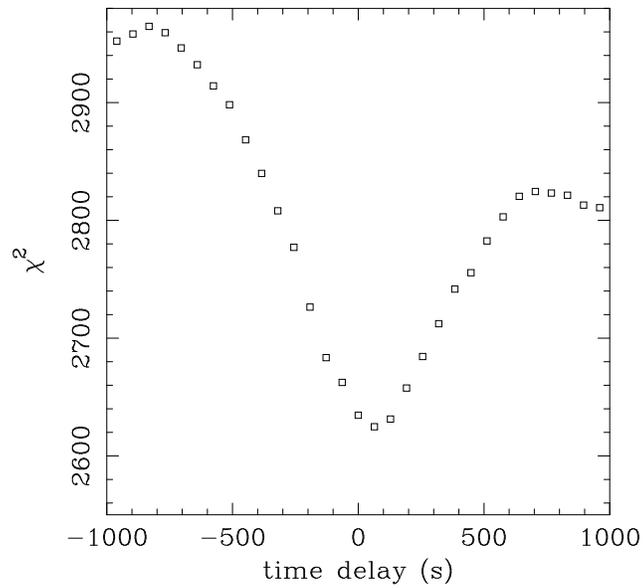}
}
\caption{The $f_{\rm tr}=1$ slice through the $\chi^2$ surface produced by
folding the 2--4\,keV light curve through the trial transfer function
$\psi_1$ and comparing with the 8--15\,keV light curve.  Note the small (1
bin) delay between the two light curves (with the harder band lagging the
softer).}
\end{figure*}

By considering the $f_{\rm tr}=1$ slice through the $\chi^2$ surface
produced with trial transfer function $\psi_1$, we can examine {\it
overall} lags and leads between energy bands.  Examining the 2--4\,keV and
5--7\,keV light curves for MCG--6-30-15 in this way, we find that the
$\chi^2$ slice possesses a minimum at zero lag --- i.e. we find no
evidence for overall time lags or leads between the continuum and line
bands down to 64\,s, the bin size of the data.  Performing the same
procedure for the 2--4\,keV and 8--15\,keV light curves reveals a one bin
offset in the position of the minimum (Fig.~4), indicating that the
8--15\,keV light curve is delayed by $\sim $50--100\,s as compared with the
2--4\,keV light curve.  

\begin{figure*}
\centerline{\psfig{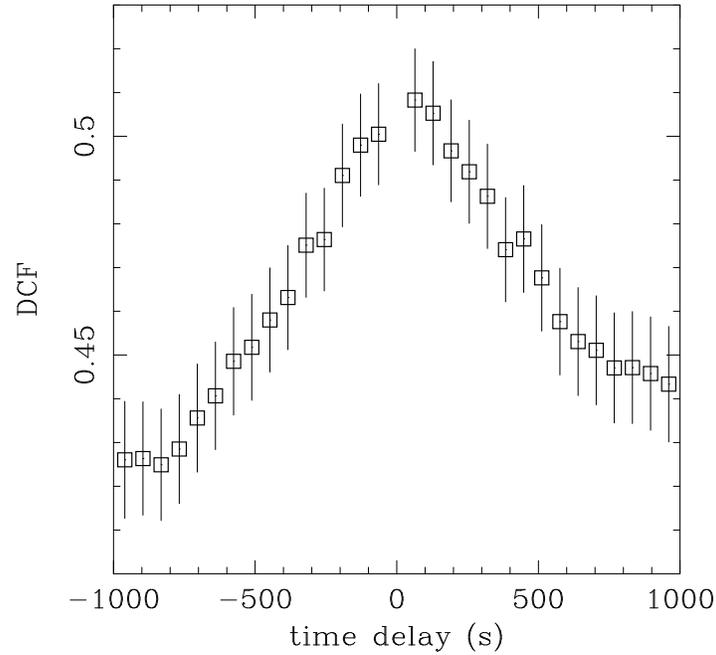}}
\caption{Discrete Correlation Function (DCF; Edelson \& Krolik 1988)
between our 2--4\,keV and 8--15\,keV light curves.  Note the asymmetry
in the DCF which validates our detection of a $\sim 50-100\s$ time lag
between these two bands using the PRH92 method.}
\end{figure*}

Lee et al. (1999b) have applied CCF methods to this {\it RXTE}
dataset. By carefully comparing with simulations, they find evidence
that the 7.5--10\,keV band lags the lower energy bands with a {\it phase
delay} of $\phi\sim 0.6$.  They also find evidence that the hard band
(10--20\,keV) lags the softer bands with a {\it time delay} similar to
that found in this work.  Figure~5 shows the DCF for our 2--4\,keV and
8--15\,keV lightcurves (this is very similar to Fig.~17 of Lee et
al. 1999b).  A small time lag of 50--100\,s between these two bands is
evident.  Thus, CCF methods and the optimal reconstruction method both
suggest a time lag of 50--100\,s between the 2--4\,keV and 8--15\,keV
bands.

\subsection{The meaning of an additive offset}

Our fitting in Section 3.2 clearly reveals the need for an additive
offset (i.e. non-zero $K$ value) between the line and continuum band
light curves.  In other words, the fractional variability about the mean
level is less in the line band that it is in the continuum band.

The spectral analysis of Lee et al. (1999b) allows this behaviour to be
understood in terms of the spectral phenomenology.  Firstly, Lee et
al. found that on timescales of ${\rm few}\times 10$\,ksec, the iron line
flux does not track the continuum flux and, instead, remains
approximately constant.  Secondly, it was found that there are flux
correlated changes in the photon index by as much as $\Delta
\Gamma\approx 0.2$ in the sense that higher flux states are softer.
Both of these spectral changes will tend to reduce variability in the
line band as compared with the (softer) continuum band.

\section{Applications to simulations}

\subsection{Constructing the simulated light curves}

\begin{figure*}
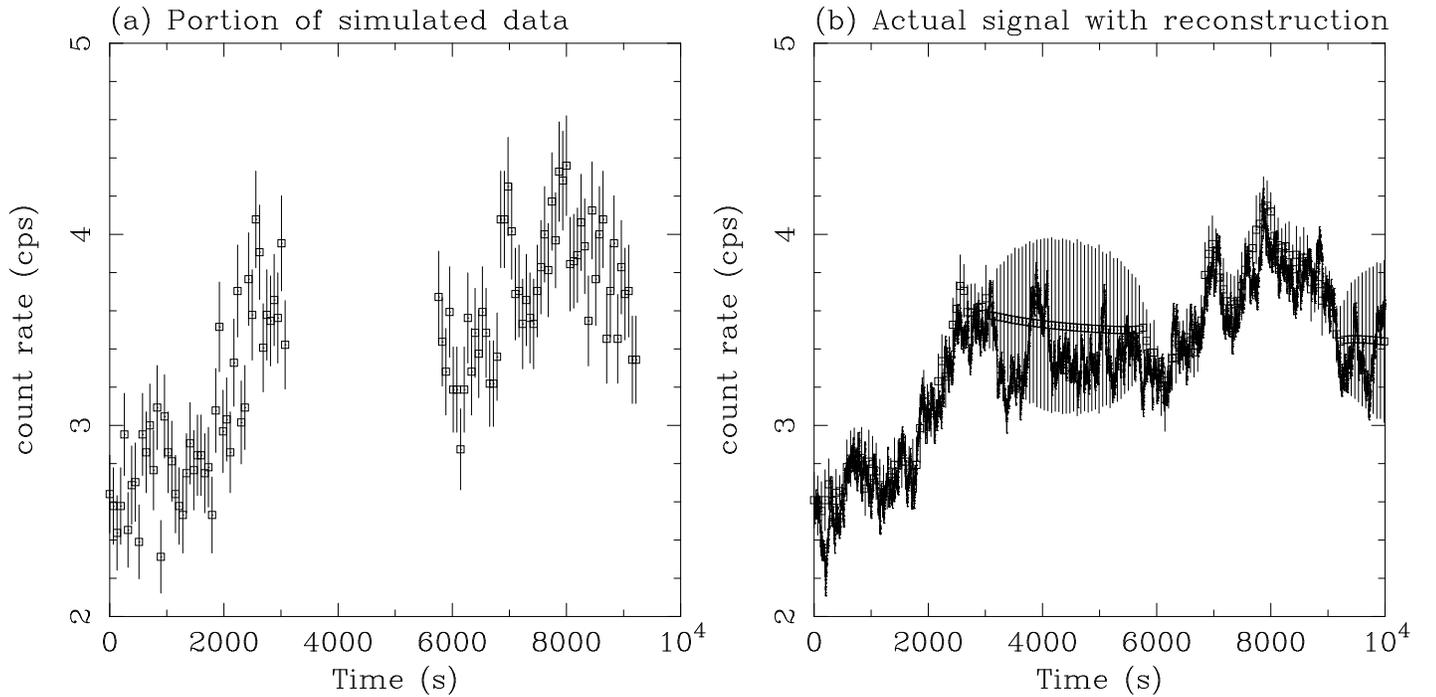

\hbox{
\psfig{figure=simulated_data.ps,width=0.5\textwidth,angle=270}
\psfig{figure=simulated_reconstruction.ps,width=0.5\textwidth,angle=270}
}
\caption{Panel (a) shows a portion of the simulated light curve described
in Section 3 of the text.  The simulated data possess 64\,s bins, and noise
is purely due to counting statistics.  The squares in panel (b) show the
results of applying the PRH92 reconstruction to the simulated data (using
$N=3000$ simulated data points.  The solid erratic line shows the `real'
simulated signal.  Note how well the reconstruction reproduces the `real'
signal during times where data exists, and brackets the signal at other
times.}
\end{figure*}

In order to assess the significance and robustness of the above results,
this section describes the application of this method to simulations.  We
tailor our simulation to match the {\it RXTE} observation MCG--6-30-15 as
much as possible.  {\it EXOSAT} showed that the high frequency fluctuations
of MCG--6-30-15 possess a power spectrum of the form $f^{-1.36}$.  We use
this power spectrum with an additional low-frequency cutoff at
$f_c=10^{-6}\Hz$:
\begin{equation}
P(f)=\left(\frac{1}{1+(f/f_c)}\right)^\alpha.
\end{equation}
We then make a (noiseless) simulated continuum light curve, $F(t)$, by
summing Fourier components of random phase between $f_{\rm min}=10^{-7}\Hz$
and $f_{\rm max}=1\Hz$, i.e.
\begin{equation}
F(t)=\int^{f_{\rm max}}_{f_{\rm min}}df\,P(f)\sin[2\pi\,f\,t-\phi(f)]
\end{equation}
where $\phi(f)$ is a uniformly randomly distributed in the range $0$ to
$2\pi$ for each distinct value of $f$.

Without loss of generality, we assume that the line band flux possesses the
same mean normalization as the continuum band light curve.  However, in
order to mimic the situation found in Section~3 as closely as possible, we
assume that there is an additive offset between the continuum band and line
band light curves as well as the convolution a transfer function.  In other
words we compute a (noiseless) line-band light curve using the expression:
\begin{equation}
b(t)=(1-f)\Lambda\,a(t)+f\int_{-\infty}^{+\infty}d\tau\Psi_{1,nzl}(\tau)a(t-\tau)+K.
\end{equation}
Here, $\Psi_{1,nzl}$ is the non-zero-lag component of our imposed simulated
transfer function for which we use a Gaussian:
\begin{equation}
\Psi_{1,nzl}={\sigma}\sqrt{\frac{2}{\pi}}\exp \left( -
\frac{(t-t_0)^2}{2\sigma^2}\right)
\end{equation}
where $f$ is the fraction of the continuum flux that is delayed, $t_0$ is
the mean time delay, and $\sigma$ is the temporal standard-width of the
smearing.  For concreteness, we set
\begin{eqnarray}
\Lambda&=&0.75\\
f&=&0.15\\
K&=&0.85\\
t_0&=&1.0\times 10^4\s\\
\sigma&=&2.0\times 10^3\s.
\end{eqnarray}
This value of $f$ is approximately the fraction of the line-band flux
which originates from the iron line, and hence this simulation crudely
mimics the effect of iron line reverberation with a $10^4\s$ time delay.
Our value of $K$ is set to be similar that found for MCG--6-30-15 above.

From these `perfect' noiseless light curves, we formed Possion sampled
noisy light curves assuming a mean count rate of 4\,cps in both the
continuum and line bands, and using 64\,s bins.  The datagap structure
of the real MCG--6-30-15 dataset was then imposed on the simulated light
curves.

We now examine our realistic, simulated, data in order to assess how well
we can detect the existence of the imposed lag and recover the properties
of $\Psi_{1,nzl}$ using our method.

\subsection{Extracting the lag from the simulations}

\begin{figure*}
\hbox{
\psfig{figure=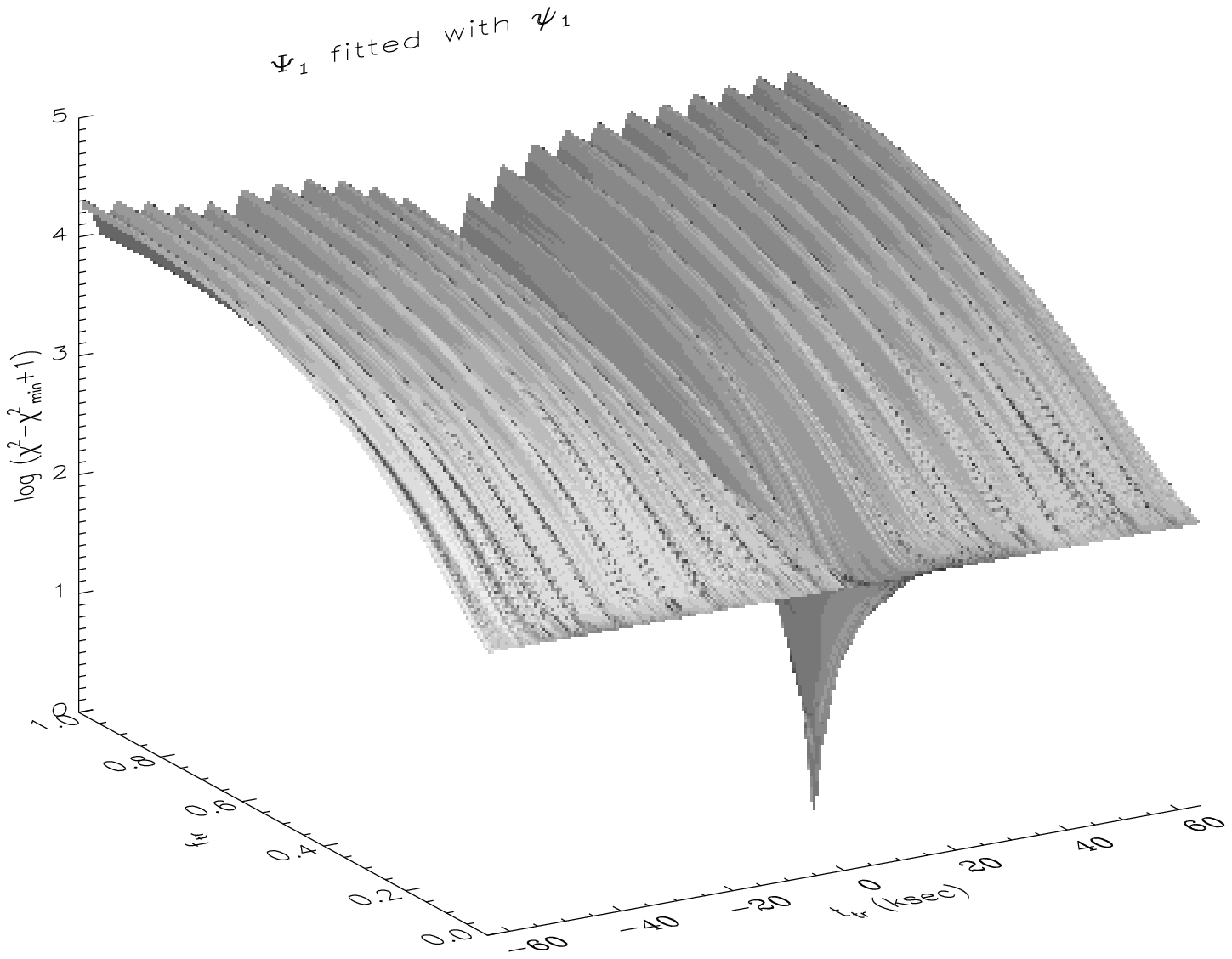,width=0.45\textwidth}
\psfig{figure=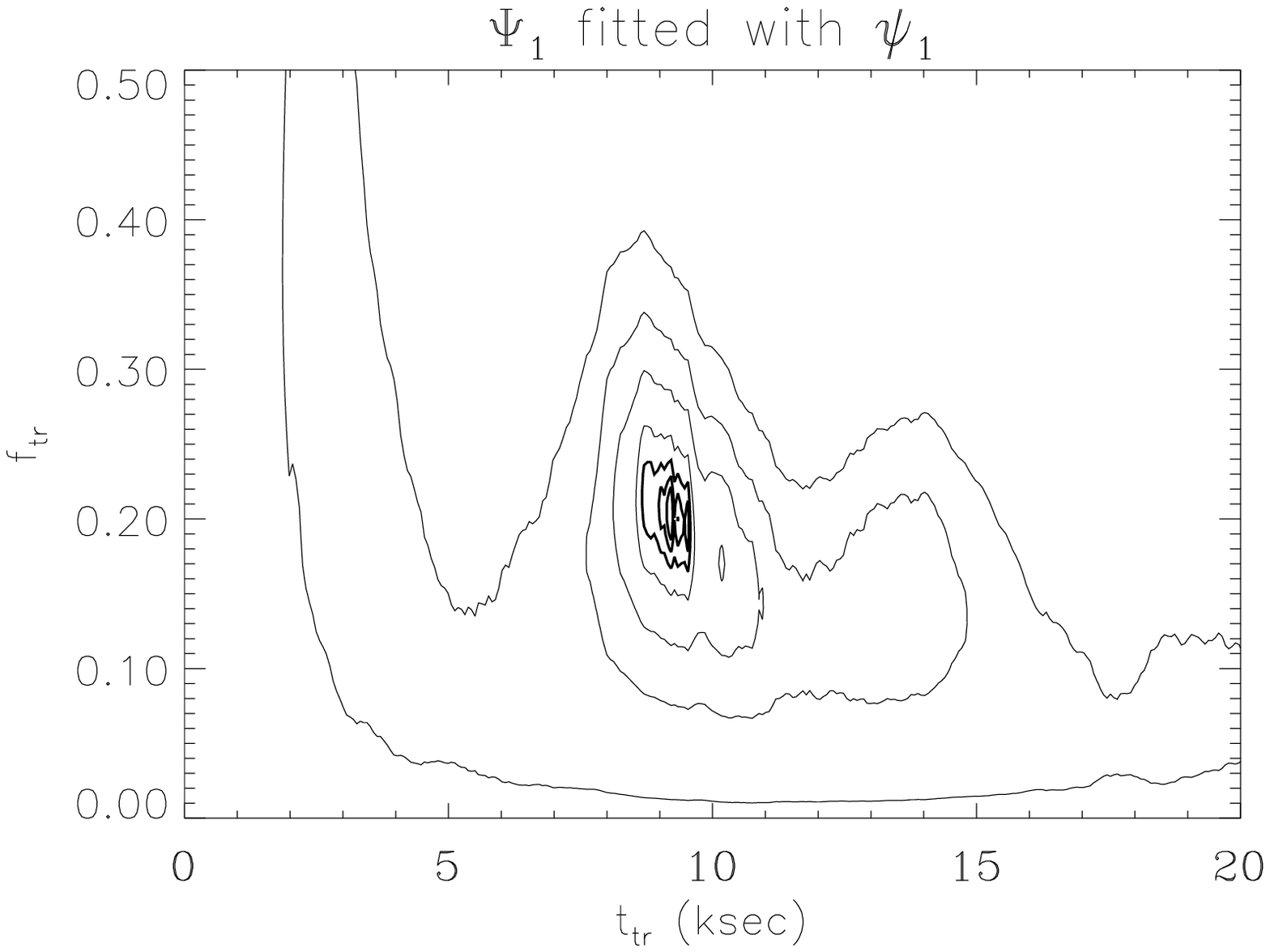,width=0.45\textwidth}
}
\hbox{
\psfig{figure=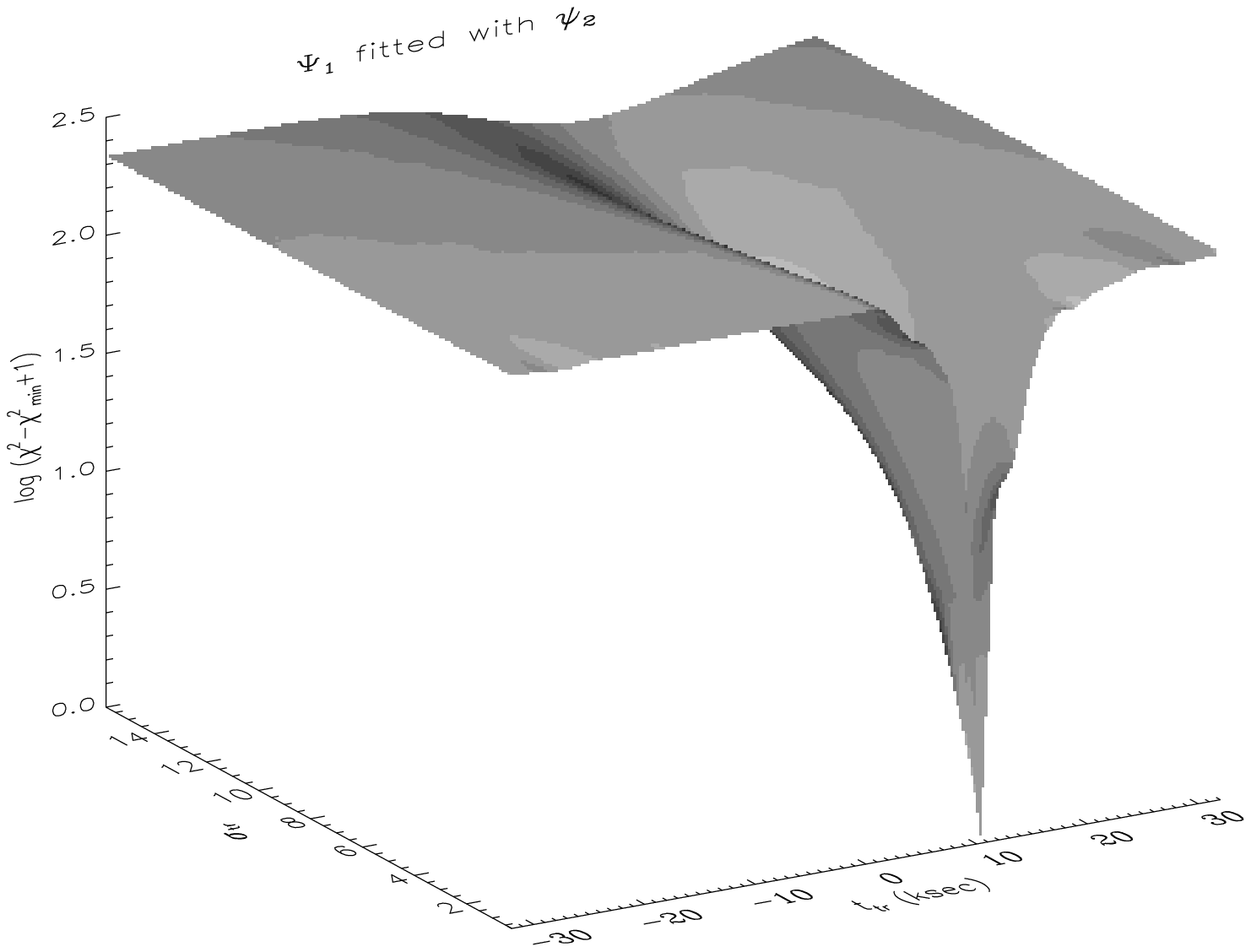,width=0.45\textwidth}
\psfig{figure=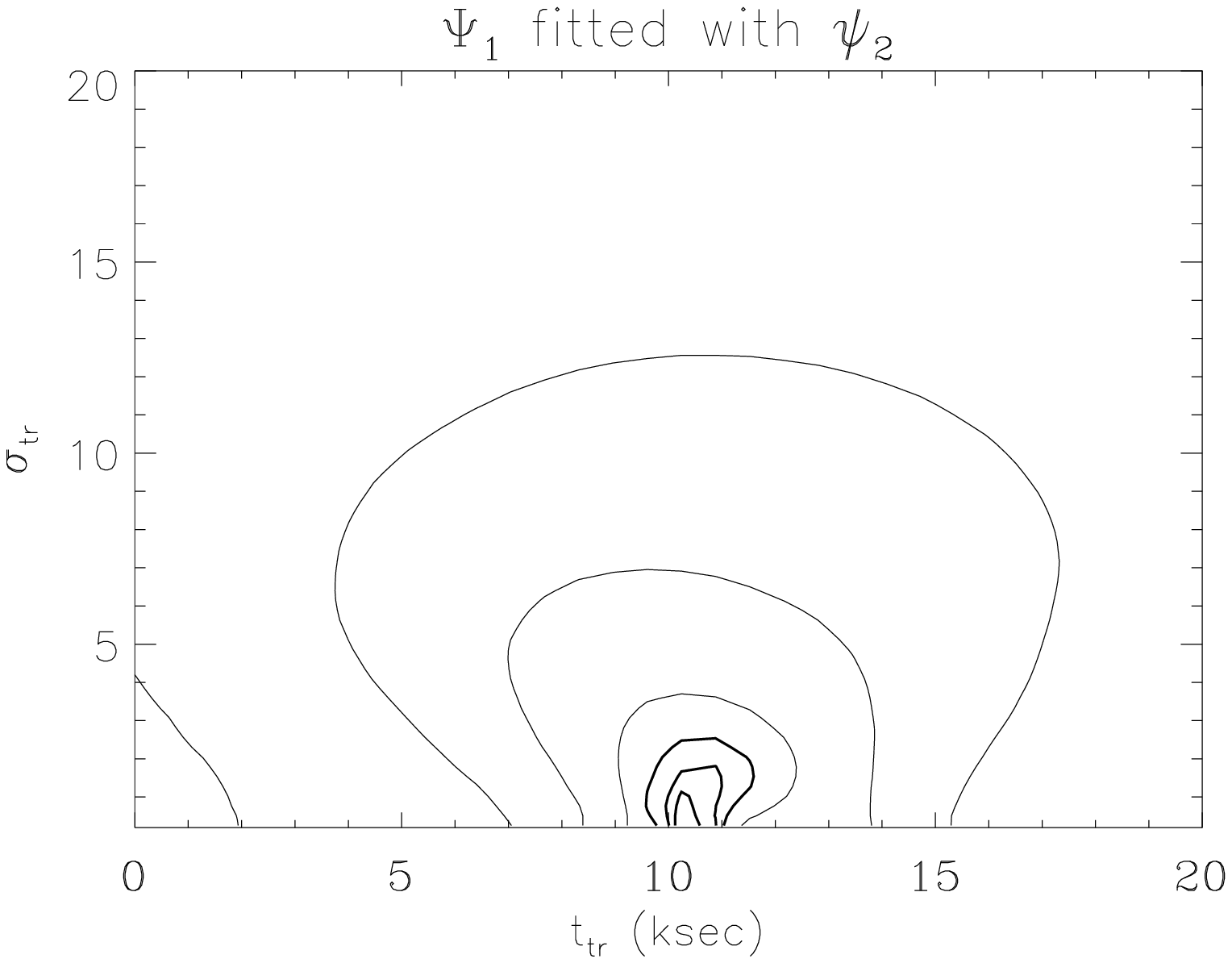,width=0.45\textwidth}
}
\caption{Results for the simulated light curves: $\chi^2$ surfaces and 
confidence contours resulting from applying trial transfer functions
$\psi_1$ and $\psi_2$ to the reconstructed continuum light curves and
comparing with the line band light curve (allowing for an additive offset
between the bands).  Surfaces are plotted using
$log_{10}(\chi^2-\chi^2_{\rm min}+1)$ as the ordinate in order to display
the topography of the region near the minimum.  Contours are shown the
following levels: $\chi^2-\chi^2_{\rm min}=2.3,4.6,9.2,20,50,100,200$.  The
first three of these contours correspond to $1-\sigma$, 90\% and 95\% for
two interesting parameters and are shown in bold.  In both cases, the
existence of a deep hole in $\chi^2$ space demonstrates that the imposed
lag has been clearly detected and its parameters recovered.}
\end{figure*}

We use the method of Section 2.1 and 2.2 to form an optimally
reconstructed, evenly-sampled continuum lightcurve.  The covariance model
used is given by eqn (13) and (14) with
\begin{eqnarray}
A&=&2.5\,{\rm cps}\\
\tau_0&=&6.23\times 10^3\s\\
\alpha_1&=&0.855\\
\alpha_2&=&1\\
\langle s\rangle&=&3.63\,{\rm cps}
\end{eqnarray}
A total of $N=3000$ simulated data points were used to form the
reconstruction which spans a simulated observation time of 400\,000\,s.  A
portion of the simulated dataset and its reconstruction are presented in
Fig.~6.  Note how well the reconstruction algorithm recovers the real
signal during the times with data, and brackets the real signal during
other times.

Figure~7 presents the $\chi^2$ surfaces and confidence contours that result
from passing the simulated light curves through the trial transfer
functions $\psi_1$ and $\psi_2$, including minimization over any additive
offset between the continuum and line band light curves.    Both trial
transfer functions clearly detect the imposed lag in so far as a a deep and
isolated hole is present in the $\chi^2$ surface at approximately the right
time delay, delay fraction and delay width.  Note that the $f_{\rm tr}$
dimension, which has been suppressed in the $\psi_2$ plots, has a value of
$f_{\rm tr}=0.16$ at the global minimum.   This demonstrates the power of
this technique for finding and characterizing subtle time lags or leads
that are present in such data. 

\section{Discussion}

In order to bring structure to the discussion that follows, we will
summarize the pertinent results from this paper.
\begin{enumerate}
\item We clearly see reduced fractional variability in the iron line band
(5--7\,keV) as compared with the continuum band (2--4\,keV).  This is the
origin of the additive offset, $K$, that was introduced in Section~3.  The
spectral fitting results of Lee et al. (1999b) suggests that this is due to
a combination of a constant iron line flux and flux correlated changes in
the photon index.
\item Our analysis finds no evidence for iron line reverberation effects.
By running a number of simulations, we find that any reverberation time
delays must be less than $\sim 500\s$ or greater than $\sim 50$\,ks.
Together with the above result, this suggests an approximately constant
iron line flux over these timescales.  Thus, we can extend the work of
Lee et al. (1999b) and infer a constant iron line on timescales down to
$0.5$\,ksec.
\item Any overall time lag between the 2--4\,keV and 5--7\,keV band is less
than $\sim 50$\,s.  However, we do find that the 8--15\,keV band is delayed
with respect to the 2--4\,keV band by 50--100\,s.  We can use this time
delay to obtain a rough size scale for the Comptonizing cloud that is
producing the hard X-rays.  Assuming a coronal temperature of $\sim
100\keV$, it takes approximately 3 inverse Compton scatterings for a
photon to be boosted between the 2--4\,keV and 8--15\,keV bands.  Thus, the
mean free path of a photon is approximately 15--30 light seconds.   This is
a lower limit on the size of the Comptonizing region.
\end{enumerate}
As we will see, this combination of facts presents problems for current
models.

\subsection{Simple reflection models}

Initially, let us discuss these results in the light of simple X-ray
reflection models (e.g. George \& Fabian 1991).  Assuming that variations
in the primary flux are not accompanied by gross changes in geometry, we
expect to observe one of two cases.  Firstly, if the light crossing time
of the fluorescing part of the disk is shorter than the timescale being
probed by the observation, an iron line with constant {\it equivalent
width} will result (i.e. the iron line flux will track the flux of the
illuminating primary X-ray source).   On the other hand, if the light
crossing time of the fluorescing disk is greater than the timescale being
probed, a constant {\it flux} line will result.

Within the context of these simple reflection models, we are forced to
conclude that the light crossing time of the fluorescing region is larger
than $\sim 50$\,ks.  Since the line is relativistically broad, most of the
fluorescence occurs in the central $r\sim 20GM/c^2$ of the disk.  Setting
the light crossing time of this region to be greater than 50\,ks gives a
black hole mass of $M_{\rm BH}\sim 2\times 10^8\Msun$.  Given such a large
black hole mass, the accretion rate must be less than 1\% of the Eddington
rate in order to produce the observed luminosity of $L_{\rm bol}\sim
10^{44}\ergps$ (Reynolds et al. 1997).  Furthermore, the size of the X-ray
emitting blobs must be small, $r_{\rm blob}/r_{\rm disk}\sim 10^{-3}$, in
order to produce the very small time delays seen between different bands.
Despite being so small, these blobs must be at large distances above and
below the accretion disk plane, or else one would still see iron line
variability as a flaring blob illuminated the patch of disk directly
beneath.

To date, there are no dynamical measurements of the black hole mass in
MCG--6-30-15, and hence such a model does not explicitly contradict any
data.  However, there are several indirect arguments that lead us to reject
the inference of a large black hole mass in MCG--6-30-15.  An independent
indicator of the black hole mass is possible by estimating the bulge mass
of the host galaxy and then applying the bulge/hole mass relationship of
Magorrian et al. (1998).  The B-band luminosity of the S0-galaxy which
hosts this Seyfert nucleus is approximately $m_{\rm B}=13.7$ (RC3
catalogue), and this is likely to be completely dominated by the bulge
since the nucleus is heavily reddened in the B-band and the galactic disk
is very weak.  Using a Hubble constant of $H_0=65\kmpspMpc$, the absolute
B-band magnitude of the bulge is then $M_{\rm B}-19$.  Using the standard
relations (Faber et al. 1997), the bulge mass is then $M_{bulge}\sim
3\times 10^9\Msun$.  Finally, applying the Magorrian et al. (1998) scaling
factor between bulge mass and black hole mass gives $M_{\rm BH}\sim
1-2\times 10^7\Msun$, an order of magnitude smaller than the black hole
mass estimate in the previous paragraph.

There are also X-ray constraints that suggest a black hole mass
significantly smaller than $10^8\Msun$.  MCG--6-30-15 has exhibited
large amplitude X-ray variability on timescales as short as 100\,s.
However, the dynamical timescale of the accretion disk where the bulk of
the energy is released is $t_{\rm dyn}\sim 10^5(M/10^8\Msun)\s$.  Thus,
if the black hole really is as massive as $M_{\rm BH}\sim 2\times
10^8\Msun$, large amplitude variability would be occurring on timescales
as short as $10^{-2}t_{\rm dyn}$.  It is difficult to conceive of
processes which would give such variability.  The final X-ray argument
against a $2\times 10^8\Msun$ black hole in MCG--6-30-15 comes from the
power spectrum derived by Lee at al. (1999b) and Nowak \& Chiang
(1999). By comparing the power spectral density (PSD) of MCG--6-30-15
with that of NGC~5548 and Cygnus X-1, they estimate that the black hole
in MCG--6-30-15 has a mass of $M_{\rm BH}\sim 10^6\Msun$.

\subsection{More complex scenarios}

Since the application of simple X-ray reflection arguments led us to deduce
an unacceptably large black hole mass, we must examine alternative avenues.
Indeed, the spectral fitting of Lee et al. (1999b) forces us to consider
complications beyond the simple reflection picture --- In their spectral
fitting, they found that the Compton reflection continuum fails to show the
expected correlation with the iron line equivalent width (in fact, they are
anti-correlated; Lee et al. 1999b).    Very similar behaviour is also
seen in NGC~5548 (Chiang et al. 1999)

Ionization of the disk surface is one of the few physical phenomenon that
can (partially) decouple the strength of the Compton reflection continuum
from the strength of the iron line.  Matt, Fabian \& Ross (1993)
demonstrated that the iron emission line is more sensitive to ionization
effects than the general form of the Compton reflection continuum.  In
other words, patches of the disk with certain (surface) ionization
parameters can produce a Compton reflection continuum without producing
appreciable iron fluorescence.

We use this fact to construct the following simple model.  Let the X-ray
flux illuminating the surface layers of the accretion disk be
\begin{equation}
F(r)\propto F_{\rm X}r^{-\beta}.
\end{equation}
A variety of X-ray source geometries give $\beta\approx 3$ at large radii,
and $\beta<3$ as one approaches the innermost parts of the disk.   Now, the
ionization parameter of at the surface of the disk is given by
\begin{equation}
\xi=\frac{4\pi F(r)}{n(r)},
\end{equation}
where $n(r)$ is the density of the surface layers of the disk.  We suppose
that $n(r)\propto r^{-\gamma}$.  Hence, we have
\begin{equation}
\xi\propto F_{\rm X}r^{\gamma-\beta}
\end{equation}
Standard disk models (Shakura \& Sunyeav 1973) give $\gamma\approx 2$ at
large radii, and $\gamma<2$ near the inner part of the disk.  Now, suppose
that there exists a critical ionization parameter $\xi_{\rm crit}$ above
which there is no iron line produced.   For reasonable values of $\beta$
and $\gamma$, this gives a critical radius $r_{\rm crit}$ within which no
iron line is produced.   The total iron line flux expected from the object
is then given by
\begin{equation}
F_{\rm line}\propto \int^{\infty}_{r_{\rm crit}}F(r)dr,
\end{equation}
which is readily manipulated to give
\begin{equation}
F_{\rm line}\propto F_{\rm X}^{(1-\gamma)/(\beta-\gamma)}.
\end{equation}
For our canonical values of $\beta$ and $\gamma$, this gives $F_{\rm
line}\propto F_{\rm X}^{-1}$.   Thus, this simple model produces an iron
line flux which is anti-correlated with the flux of the illuminating
source.   Provided a strong Compton reflection continuum can still
originate from the ionized portions of the disk, this type of picture may
explain the spectral behavior that we observe.   

One simple prediction of this model is that the velocity width of the
line profile gets smaller as the continuum flux increases (due to an
outward migration in the inner radius of the line emitting region).  Of
course, the toy model presented above only captures the crudest aspects
of the problem.  Fully self-consistent ionized reflection models must be
calculated (taking into account the vertical structure of the disk;
e.g. see Nayakshin, Kazanas \& Kallman 1999) and compared with the data
in order to test whether the picture sketched here is reasonable or not.

Even if global, flux-correlated changes in the ionization of the disk
surface are responsible for the observed spectral changes, we would still
expect reverberation signatures on short timescales.  We have set upper
limits of $\sim 500$\,s on the timescale of any reverberation delay.  If
the black hole mass is $M_{\rm BH}\sim 1\times 10^7\Msun$, the light
crossing time of the iron line producing region is $\sim 2000$\,s, and
hence we need to infer a disk-hugging corona (with $h/r\sim 0.3$) in order
to be compatible with the reverberation limits.  If, instead, the black
hole is $M_{\rm BH}\sim 1\times 10^6\Msun$, the light crossing time of the
entire line producing region is only $200\s$ and so the X-ray source
geometry is unconstrained by our reverberation limits.  The corresponding
Eddington ratios are $\sim 10\%$ and $\sim 100\%$ for black hole masses of
$10^7\Msun$ and $10^6\Msun$ respectively.

\section{Conclusions}

In this paper, we have used an interpolation method based upon that of
PRH92 to search for temporal lags and leads between the 2--4\,keV,
5--7\,keV and 8--15\,keV bands in a long {\it RXTE} observation of the
bright Seyfert 1 galaxy MCG--6-30-15.  In essence, we use the PRH92 method
to compute an optimal reconstruction of the 2--4\,keV light curve in which
the datagaps are interpolated across.   We then fold this reconstructed
light curve through trial transfer functions and compare with data from the
other bands in a $\chi^2$ sense.  

Our search for lags and leads was tailored to find reverberation effects in
the iron line which is thought to originate from the innermost regions of
the black hole accretion disk.  We find no evidence for any reverberation,
and rule out reverberation delays in the range $0.5-50$\,ksec.  We
can extend the conclusions of Lee et al. (1999b), and infer that the
iron line possesses a constant flux on timescales on timescales as short
as 500\,s.   We also find that the hard band (8--15\,keV) is
delayed by 50--100\,s relative to the 2--4\,keV band.

We attempt to put these various results together into a coherent picture
for this object.  The constancy of the iron line flux leads one to consider
large black hole masses (in excess of $10^8\Msun$).  However, such a large
mass is found to be unacceptable from the standpoint of both X-ray
variability constraints, and constraints based on the mass of the galactic
bulge.  Indeed, using the bulge/hole scaling factor of Magorrian et
al. (1998), we estimate that the hole has a mass of $M_{\rm BH}\sim
1-2\times 10^7\Msun$.  Given that this is a more reasonable mass estimate,
some mechanism beyond the simple X-ray reflection model must be invoked to
explain the temporal variability of the iron line and Compton reflection
continuum.  We suggest that flux correlated changes in the average
ionization state of the surface layers of the accretion disk may be such a
mechanism.  While we support this suggestion with a toy model, the
plausibility if this suggestion can only be assessed once detailed
modeling has been performed.

\section*{Acknowledgments}

I thank James Chiang, Rick Edelson, Andrew Hamilton, Julia Lee and Mike
Nowak for insightful discussions throughout the course of this work.
CSR thanks support from Hubble Fellowship grant HF-01113.01-98A.  This
grant was awarded by the Space Telescope Institute, which is operated by
the Association of Universities for Research in Astronomy, Inc., for
NASA under contract NAS 5-26555.  CSR also thanks partial support from
NASA under LTSA grant NAG5-6337.  This work has made use of data
obtained through the High Energy Astrophysics Science Archive Research
Center (HEASARC) Online Service, provided by the NASA Goddard Space
Flight Center.

\end{document}